\documentclass[twocolumn,showpacs,prd]{revtex4}
\usepackage{dcolumn}
\usepackage{bm}
\usepackage[normalem]{ulem}
\usepackage{graphicx}
\begin{document}

\title{Analysis of an attempt at detection of neutrons produced in a plasma discharge electrolytic cell }
\author{A. Widom and J. Swain}
\affiliation{Physics Department, Northeastern University, Boston MA 02115}
\author{Y. N. Srivastava}
\affiliation{Physics Department \& INFN, University of Perugia, Perugia IT}
\author{D. Cirillo}
\affiliation{1.21GW Research Laboratory, Caserta, IT}

\begin{abstract}
R. Faccini {\it et al.} \cite{Faccini:2013} have attempted a replication of an earlier experiment 
by D. Cirillo {\it et al.} \cite{Cirillo:2012} in which neutrons [as well as nuclear transmutations] 
were observed in a modified Mizuno cell. No neutron production is observed
in the recent experiment \cite{Faccini:2013} and no evidence for microwave radiation or 
nuclear transmutations are reported. A careful analysis shows major technical differences 
in the two experiments and we explore the underlying reasons for the lack of any nuclear 
activity in the newer experiment. 
\end{abstract}


\maketitle

\section{Introduction \label{com}}

For over a decade, concentrated experimental efforts have been dedicated to perfecting
conditions appropriate for igniting nuclear activity in modified electrolytic plasma Mizuno cells. 
Detection of a distinct neutron flux using modified CR-39 neutron counters in a host of specially
prepared electrolytic cells  was presented in \cite{Cirillo:2012} and nuclear transmutations 
were also observed in a series of such experiments. 

Hence, the lack of any nuclear activity in \cite{Faccini:2013} done with a meagre set
of three runs, yet purporting to replicate the earlier exhaustive sets of experiments culminating 
in \cite{Cirillo:2012}, may at first sight appear contradictory. But we shall show 
in this paper that there are substantial technical differences both in the preparation of the 
conditions within the electrolytic cell as well as in the neutron measurement technique to 
render the two experiments strikingly apart. Ignition of a nuclear reaction occurs only under specific
experimental conditions and hence they need to be followed. From the details of the set up 
described in \cite{Faccini:2013}, it would appear as if the newer experiment 
was designed to suppress the onset of any nuclear activity.   
 
In Sec.(\ref{cathode}), the differences in cathode preparation, its cover and its positioning
relative to the anode  in the two experiments are discussed. Plasma conditions between 
the two experiments are reviewed in Sec.(\ref{plasma}). The choice and physical positioning
of their neutron detectors are considered in Sec.(\ref{neutron}). In the concluding Sec.(\ref{conc}), 
we briefly indicate how the drastically modified conditions in \cite{Faccini:2013} are 
responsible for their null results.

\section{Preparation and positioning of the cathode \& the anode \label{cathode}}
Surface dynamics is notoriously difficult to control and hence specific conditions must be 
satisfied for a proper comparison between two experimental set ups.

Workers experienced in the workings of a Mizuno type cell know how important is
the preparation and positioning of the cathode surface. For the purpose at hand (ignition
of nuclear activity), the cathode surface needs to be cracked, with sharp points 
but certainly not a regular surface. The theoretical reason for a ``rough'' versus a smooth surface 
is that near a sharp or irregular point the electric field is much larger thus leading to a larger  
$\gamma$ factor for the excited electron mass \cite{Cirillo:2012}\cite{Widom:2013}:
\begin{equation}
\label{1}
m^* = \gamma m,
\end{equation}
and $\gamma \geq 2.5$ for neutron production. Sharper points on the electrode surface 
also give rise to hot spots on the surface where extensive nuclear activity can occur. Absorption 
of a significant number of protons on the metal surface produces a non-smooth surface. No details are 
available in \cite{Faccini:2013} about this point.

The inner diameter of the tube covering the cathode is an important parameter. A given 
diameter shall lead to a specific plasma condition and different choices would lead to different 
plasmas. We stress the obvious point that different plasmas give different results. Some 
fundamental requirements have to be carefully considered for a given set up and the inner 
diameter has to be appropriately chosen. No discussion about this matter has been provided 
in \cite{Faccini:2013}.

The positioning of the anode and other features such as its mesh and dimensions are important 
conditions to obtain a given plasma. There is no information given on this subject in \cite{Faccini:2013}.

Another important discernible difference between the two experiments is the following. In \cite{Faccini:2013},
there is a thick quartz wall between the cathode and the anode. Its presence modifies the experimental
set up considerably. There is no such wall separating the two electrodes in \cite{Cirillo:2012}. We shall return to
this point in the next section.

\section{ Plasma conditions \label{plasma}}
The cathode cover used in \cite{Faccini:2013} is quartz and alumina in \cite{Cirillo:2012}. This cover serves
to expose a proper surface for the plasma and for driving the electronic paths (the electrolytic current)
in a region far from the plasma (an area of higher quiescence). It has to provide a correct passage for 
the electric current essential for
a sustained stable plasma. If the conditions are not met, the plasma remains unstable. Hence, the inner
diameter of this pipe cover is a sensitive parameter that ultimately determines whether an experiment 
is of a good quality.  

The quality referred to in the last paragraph concerns the stability of the plasma in the cell. It is 
discerned quite reliably through the voltage-current ($V-I$) characteristic. It is well known 
from Mizuno cell type experiments under discussion, such as in \cite{Cirillo:2012}, that stable glow discharges with
stable currents are achieved for voltages in a given range particular for each set up. We show in Fig.(1) 
the voltage $V$ and the current $I$ as a function of time for one of the cells in \cite{Cirillo:2012}. For this case,
the appropriate range is between about ($210 \pm 20$) Volts. The region of voltage 
below this range for this cell would be unsuitable. 

\begin{figure}
\scalebox {0.6}{\includegraphics{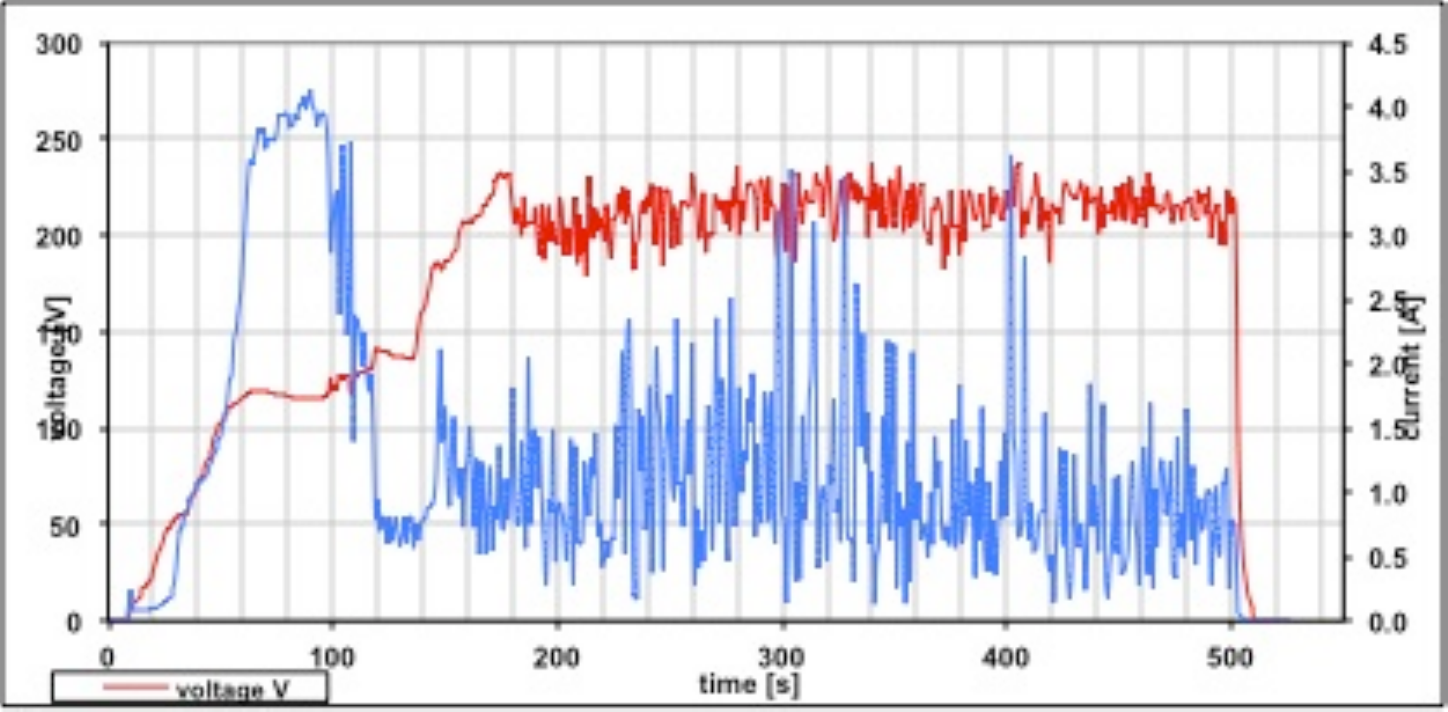}}
\caption{The Voltage $V$ and the current $I$ are shown as a function of time.} 
\label{fig1}
\end{figure}

No $V-I$ characteristic is shown in \cite{Faccini:2013}. However, the trend of
CR-39 neutron flux shown there for one of the runs [Run 1A] is $(280\pm32) {\mathrm{n/cm^2sec}}$ for ($150 - 200$) Volts
whereas it {\it decreases} to $(166\pm 32) {\mathrm {n/cm^2sec}}.$ for the larger range ($150 - 300$) Volts in 
another run [Run1D]. Such an odd and unphysical result confirms our assertion that at best their current and 
hence their plasma cannot have been very stable. This is a serious lacuna in the Rome 
experiment \cite{Faccini:2013}. 

In  Fig.(1) of the Rome paper \cite{Faccini:2013}, the colour of the plasma shown due to potassium excitation
appears violet interspersed with cloudy light around the cathode and quite likely the temperature reached is 
insufficient for the purpose at hand. By contrast, there is a steady glowing white plasma in \cite{Cirillo:2012}.  

\section{Neutron detection \label{neutron}}
Next we turn to an analysis of the neutron detection techniques employed in the two experiments. 

Low energy neutron production on an electrode surface (via surface plasmon polaritons for example) 
is theoretically expected to occur through random bursts, it is not a continuous production process. It is 
for this reason that neutron detection through indium detectors that are calibrated from known continuous 
neutron sources is not a reliable device\cite{indium}. In fact, no neutrons were detected through indium detectors in
\cite{Cirillo:2012}. Not surprisingly a similar result was found also in \cite{Faccini:2013}.

On the other hand, there is a sharp difference between the results of the two experiments with CR-39
detectors. The method based on CR-39 +  boron can provide a  measurement of the plasma-generated 
neutron flux albeit with a low efficiency. Hence, this method can and has been used to obtain clear evidence 
of thermal neutron generation, whereas other methods might fail ({\it i.e.} conventional sample saturation, 
electromagnetic detectors). 


In the Rome experiment \cite{Faccini:2013},  the CR-39 etching treatment had a duration of around  
90 minutes at a temperature of $70^{o}$C with 6.25 N KOH. In the Naples experiment \cite{Cirillo:2012}, 
such a process had a duration of around 5 hours at $63^{o}$C with 6.2\ M NaOH. Considerable time
period difference in the etching process can cause a substantial change in the counting of tracks.

When CR-39 is covered with boron, the granularity of the boron layer used in \cite{Faccini:2013} is unknown 
whereas for the experiment \cite{Cirillo:2012} the granularity of the boron powder is an important parameter 
for detection efficiency. This parameter determines the ``mean free path'' of the emitted alpha particles.

In the Rome experiment \cite{Faccini:2013}, CR-39 + boron gave no signal when exposed to a known thermal 
neutron flux. We quote from \cite{Faccini:2013}: {\it ``At the same time, CR-39 detectors covered with a thick
boron coating were exposed to the calibration flux, but no significant signal was observed''.} 

This is a strong clue towards substantial differences in the detector assembly of the two experiments. 
Apparently, the CR-39 + boron detector in \cite{Faccini:2013} did not work properly. 

By contrast, we show in Fig.(2) the calibration of the CR-39 + boron detector. At the top of each sample,
one can find the value of the measured number of tracks$/{\mathrm{mm}}^2$.
 
\begin{figure}
\scalebox {0.6}{\includegraphics{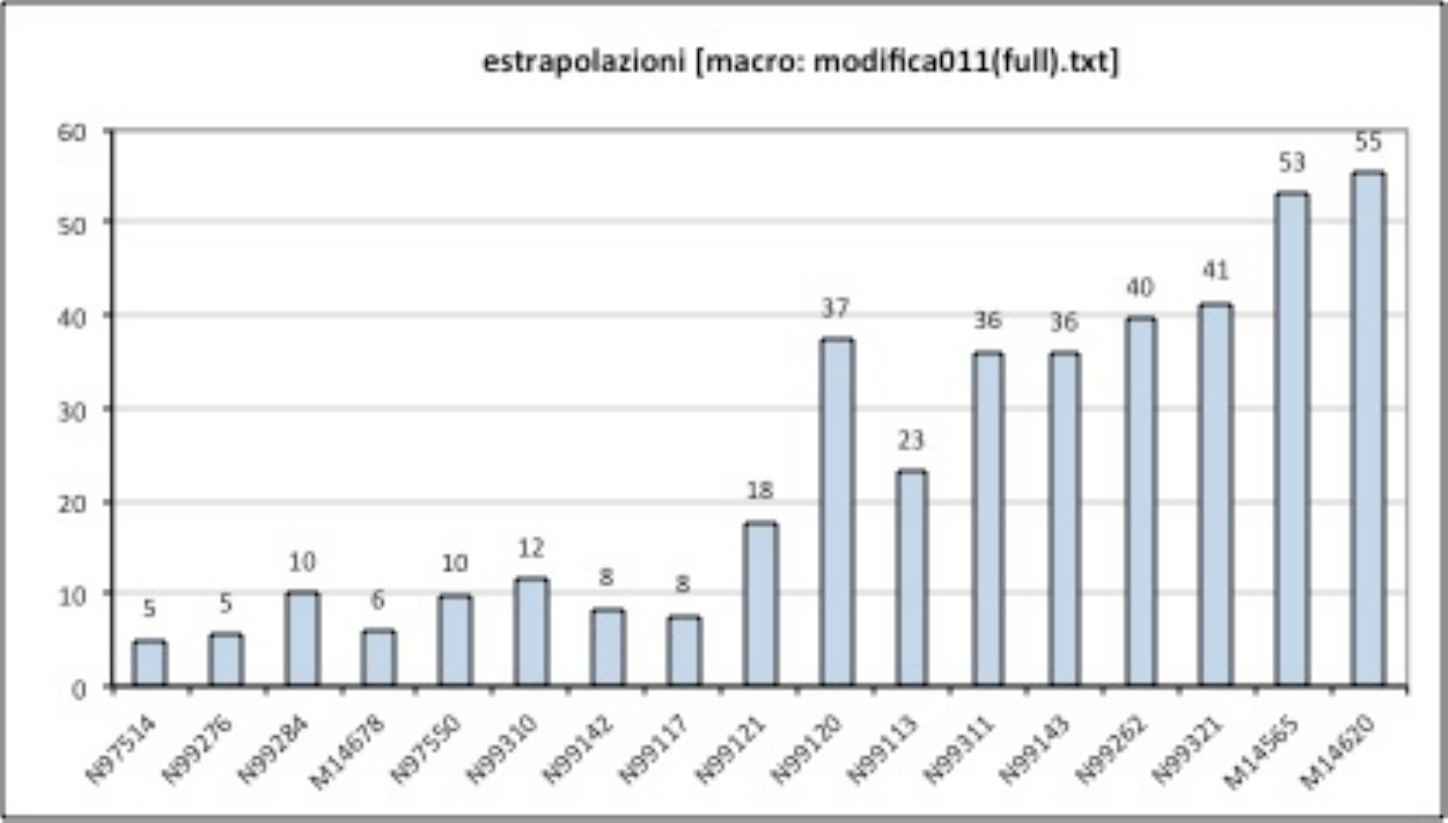}}
\caption{Calibration of the CR-39 neutron detector.
This figure shows the track density for each calibration sample P/N ({\it i.e.} the sample N97514 shows 5 tracks/mm$^2$). 
The samples in this plot were exposed to a well known thermal neutron flux at ENEA.} 
\label{fig2}
\end{figure}
In Fig.(3) we show the measured track density through the apparatus\cite{Cirillo:2012}.
In this figure, the horizontal plots (CR-39 Sample 1, 2, 3, 4, 5) show a given sample exposed for a given time to the
known thermal neutron flux so that we know that the track density measured by CR-39 sample 1, for
example, is similar to the track density exposed to the standard neutron flux for ($50 \div 60$) minutes.
Hence, comparisons can be made between known and plasma track densities since the standard
one depends on the time of exposure.

\begin{figure}
\scalebox {0.6}{\includegraphics{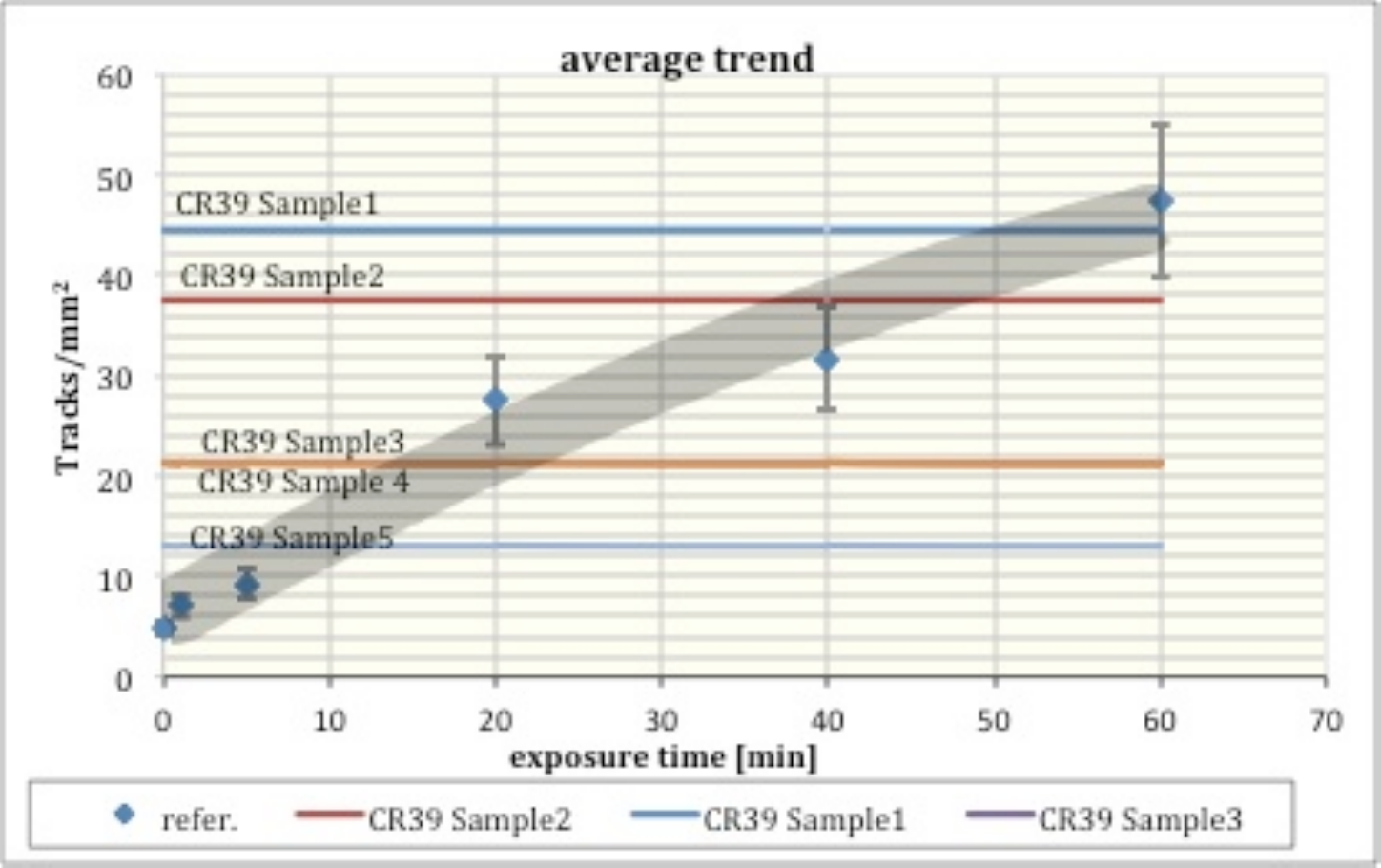}}
\caption{The number of tracks/mm$^2$  is shown as a function of the time of exposure (in minutes). 
See text for explanations and more details about this plot.} 
\label{fig3}
\end{figure}


\section{Conclusion \label{conc}}
Our analysis identifies several important differences between the two experiments both in 
the preparation of the cell, the characteristics of the plasma, the choice of the neutron detectors {\it etc.}, so much so
that it would be inappropriate to call \cite{Faccini:2013} a replica of \cite{Cirillo:2012}. On this basis, it is also
safe to conclude that all of the three runs described in \cite{Faccini:2013} were geared to produce no nuclear
activity and hence no neutrons, no microwave radiation and no nuclear transmutations \cite{Violante:2013}. 
Hence, this experiment fails to provide any valid criticism against substantial neutron production and nuclear 
transmutations found through myriads of experiments with stable plasma as in \cite{Cirillo:2012}.

Notwithstanding uncertainties inherent in the neutron detection processes as discussed in \cite{Cirillo:2012} and in 
previous experiments, the Rome experiment \cite{Faccini:2013} is not a falsification of the nuclear activity 
results found in \cite{Cirillo:2012}. A true falsification of an experimental result requires an accurate and exact 
replication of it by another experiment. The experiment described in \cite{Faccini:2013} does not do this correctly.  
     
\section{Acknowledgement}
YS would like to thank Dr. V. Violante for useful correspondence regarding changes in his conclusions about his 
past experiment with surface plasmon polaritons and the experiment described in \cite{Faccini:2013}.

\end{document}